\documentclass[aps,prb,reprint,nofootinbib,superscriptaddress,longbibliography]{revtex4-2}

\usepackage{amssymb}
\usepackage{multirow}
\usepackage{graphicx}
\usepackage{amsmath}
\usepackage{color}
\usepackage{mathrsfs}
\usepackage{indentfirst}
\usepackage{textcomp}
\usepackage{comment}
\usepackage{mathtools}
\usepackage{hyperref}
\usepackage{soul}            
\usepackage{braket}

\begin{document}
\title{On the precise boundary of the N\'eel antiferromagnetic phase in the Shastry-Sutherland model}

\author{Rongyi Lv}
\affiliation{Key Laboratory of Artificial Structures and Quantum Control (Ministry of Education),  School of Physics and Astronomy, Shanghai Jiao Tong University, Shanghai 200240, China}

\author{Xiangjian Qian}
\affiliation{Key Laboratory of Artificial Structures and Quantum Control (Ministry of Education),  School of Physics and Astronomy, Shanghai Jiao Tong University, Shanghai 200240, China}
\affiliation{Tsung-Dao Lee Institute, Shanghai Jiao Tong University, Shanghai 200240, China}

\author{Mingpu Qin} \thanks{qinmingpu@sjtu.edu.cn}
\affiliation{Key Laboratory of Artificial Structures and Quantum Control (Ministry of Education),  School of Physics and Astronomy, Shanghai Jiao Tong University, Shanghai 200240, China}

\affiliation{Hefei National Laboratory, Hefei 230088, China}

\date{\today}


\begin{abstract}
    The ground state phase diagram of the Shastry-Sutherland model remains controversial. In particular, the nature of the transition between the plaquette valence bond solid (PVBS) phase and the N\'eel antiferromagnetic (AFM) phase is still under debate. While some studies suggest a direct transition between the two phases, others propose the existence of a narrow intermediate quantum spin liquid regime. To address this issue, we perform large-scale Density Matrix Renormalization Group calculations to determine the N\'eel AFM phase boundary in the Shastry-Sutherland model. By applying staggered boundary pinning fields and directly measuring the bulk staggered magnetization, and through systematic extrapolations with respect to truncation error as well as careful finite-size scaling, we find that the staggered magnetization remains finite across the controversial region $0.78 \lesssim J_{1}/J_{2} \leq 0.81$, which was suggested to host a quantum spin liquid phase in some studies. Together with the previously established consensus placing the PVBS phase boundary at $J_{1}/J_{2} \approx 0.78$, these results provide evidence against the existence of an intermediate quantum spin liquid phase and support a direct PVBS–N\'eel AFM transition in the Shastry-Sutherland model.
\end{abstract}

\maketitle

{\em Introduction --}
Frustrated quantum magnets provide a fertile platform for exploring exotic quantum phases and phase transitions beyond the conventional Landau-Ginzburg paradigm. Strong frustration and quantum fluctuations may give rise to long-range entangled states such as quantum spin liquids (QSLs)~\cite{anderson1973resonating, doi:10.1126/science.235.4793.1196, balents2010spin, savary2017quantum,RevModPhys.89.025003}, which lack long-range order even at zero temperature and fall outside the traditional symmetry-breaking framework. Elucidating the nature of such exotic phases is of fundamental importance for the study of strongly correlated quantum magnets. 

\begin{figure*}[t]
    \includegraphics[width=160mm]{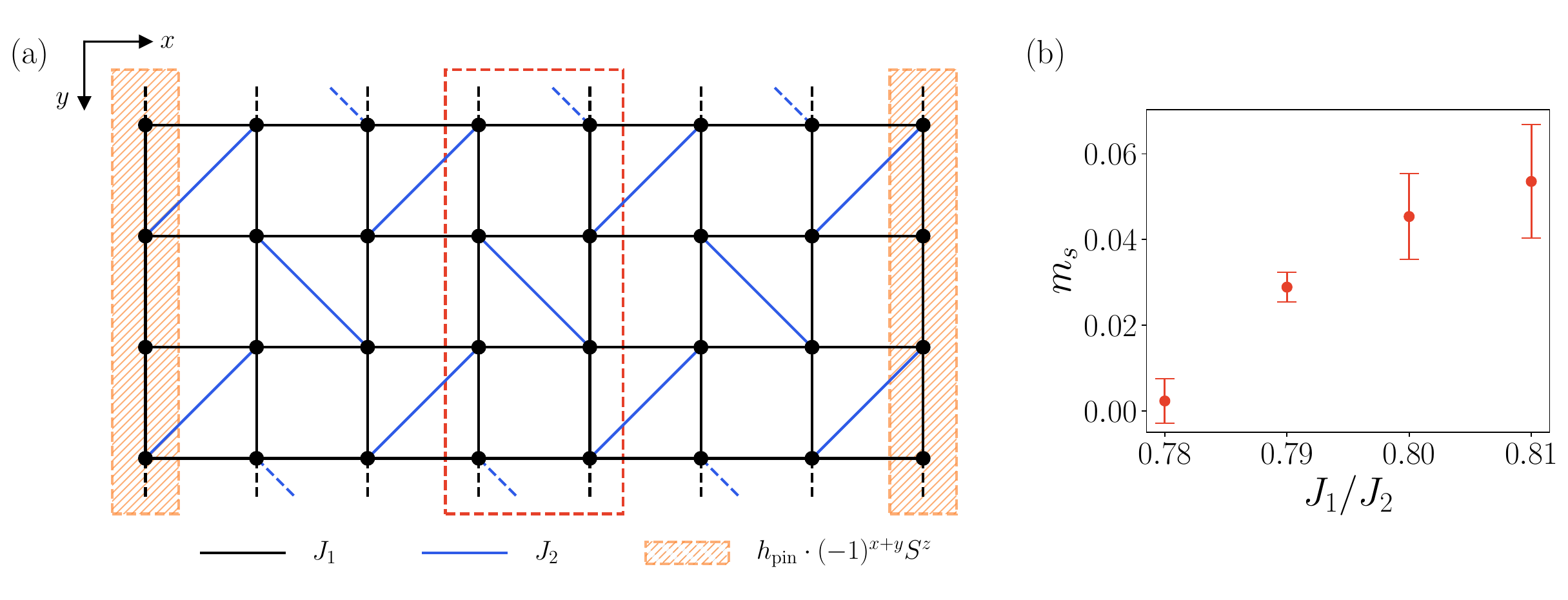}
    \caption{Shastry-Sutherland model and the results for staggered magnetization.
    (a) Shastry-Sutherland model on a $4 \times 8$ cylinder with open boundary conditions in the $x$ direction and periodic boundary conditions in the $y$ direction. The black and blue lines represent the $J_{1}$ and $J_{2}$ interactions, respectively. We set $J_2=1$ throughout this work. A pinning field along the $z$ direction with strength $h_{\mathrm{pin}}$ is applied at the open boundaries (orange hatched regions). The staggered magnetization $m_{s}$ is measured on the two central columns (red dashed box).
    (b) Staggered magnetization $m_{s}$ as a function of $J_{1}/J_{2}$, measured with a pinning field ($h_{\mathrm{pin}} = 0.5$) and extrapolated to the thermodynamic limit using results from multiple cylinders with different aspect ratios (details in Fig.~\ref{pin_check} and Fig.~\ref{Ms_center_SSM_combined}). The extrapolated $m_{s}$ places the lower boundary of the N\'eel AFM phase at $J_{1}/J_{2} \approx 0.78$.}
    \label{plaquette}
\end{figure*}

Among various frustrated quantum magnet models, the Shastry-Sutherland model has attracted significant attention over the past few decades. It is also believed to capture the essential physics of the quasi-two-dimensional material $\text{SrCu}_{2}(\text{BO}_{3})_{2}$~\cite{SRIRAMSHASTRY19811069, PhysRevLett.82.3168, PhysRevLett.84.4461, doi:10.1126/science.1075045, PhysRevB.100.140413, jimenez2021quantum, doi:10.1126/science.adc9487, cui2025plaquettesingletphasesemergentso5, guo2025deconfined}. The Hamiltonian of the Shastry-Sutherland model is given by
\begin{equation}
    H = J_{1} \sum_{\langle i, j \rangle} \mathbf{S}_{i} \cdot \mathbf{S}_{j} + J_{2} \sum_{\langle\langle i, j \rangle\rangle} \mathbf{S}_{i} \cdot \mathbf{S}_{j},
\end{equation}
where $\mathbf{S}_{i}$ is the spin-1/2 operator at site $i$. The antiferromagnetic coupling constants $J_{1}$ and $J_{2}$ represent the nearest-neighbor ($\langle i, j \rangle$) and next-nearest-neighbor ($\langle \langle i, j \rangle \rangle$) interactions, respectively. As shown in Fig.~\ref{plaquette} (a), the nearest-neighbor bonds are represented as black pairs and the next-nearest-neighbor bonds are represented as blue pairs. Different from the $J_1$-$J_2$ Heisenberg model on the square lattice, the Shastry-Sutherland model contains only part of the next-nearest-neighbor interactions. But same as in the $J_1$-$J_2$ Heisenberg model, the competition between $J_1$ and $J_2$ couplings in the Shastry-Sutherland model induces strong frustration, resulting in a diverse and complex ground state phase diagram. It is widely accepted that for $J_{1}/J_{2} \lesssim 0.68$, the system resides in a dimer singlet phase, while for $J_{1}/J_{2} \gtrsim 0.82$, it enters the N\'eel antiferromagnetic (AFM) phase. Between these two phases, an intermediate plaquette valence bond solid (PVBS) phase in the range of $ 0.68 \lesssim J_{1}/J_{2} \lesssim 0.78$ has been established in previous studies~\cite{PhysRevLett.84.4461, PhysRevB.87.115144, PhysRevX.9.041037, FRG2022, PhysRevB.105.L060409, Anders2022CPL, PhysRevB.107.L220408, PhysRevLett.133.026502, chen2025spinexcitationsshastrysutherlandmodel, Viteritti2025, Qian2025, Z2diracspinliquid2026, LvchengChen2026, Corboz2025}.

However, the nature of the phase transition between the PVBS and AFM phases remains under active debate. On the one hand, several numerical studies suggest a direct transition between the two ordered phases. Early density matrix renormalization group (DMRG) results proposed a deconfined quantum critical point (DQCP) with emergent O(4) symmetry at $J_1/J_2 = 0.77$~\cite{PhysRevX.9.041037}, while recent finite projected entangled pair state (PEPS) calculations also support a DQCP scenario, but locating the transition at $J_1/J_2 = 0.828$~\cite{PhysRevLett.133.026502}. Alternatively, studies based on projected entangled simplex state~\cite{PhysRevB.107.L220408} and DMRG~\cite{Qian2025} support a weakly first-order transition at $J_1/J_2 \approx 0.79$. On the other hand, multiple numerical investigations, including functional renormalization group~\cite{FRG2022}, DMRG~\cite{PhysRevB.105.L060409}, neural quantum states~\cite{Viteritti2025}, and very recent infinite PEPS (iPEPS)~\cite{Corboz2025}, have suggested the existence of a narrow intermediate QSL phase in the range of $0.78 \lesssim J_1/J_2 \lesssim 0.82$. Despite these different scenarios, previous studies have established a broad consensus that the PVBS phase terminates at $J_1/J_2 \approx 0.78$~\cite{PhysRevX.9.041037, FRG2022, PhysRevB.105.L060409, Anders2022CPL, PhysRevB.107.L220408, chen2025spinexcitationsshastrysutherlandmodel, Viteritti2025, Qian2025, Corboz2025}. Therefore, the key unresolved question is whether N\'eel AFM order develops at the boundary of the PVBS phase, corresponding to a direct PVBS--N\'eel AFM transition, or only at a larger value of $J_{1}/J_{2}$, leaving an intervening magnetically disordered region that may host a QSL phase.


A central difficulty in resolving this issue lies in the reliable determination of the AFM order parameter near the phase boundary. In finite-size simulations, preserving SU(2) spin symmetry precludes a direct measurement of the staggered magnetization $m_{s}$ from the local spin expectation values. Approaches based on finite-size scaling of spin correlations effectively compute $m_{s}^{2}$ rather than $m_{s}$ itself and can therefore suffer from significant uncertainties when the magnetic order parameter is very small~\cite{Sandvik_2010, Stoudenmire_2012, pinningtheorder_prx2013, Qian2025}, especially in the narrow parameter region where different phases may compete. In such cases, distinguishing a genuinely disordered phase from a weakly ordered AFM phase is challenging.

In this work, we revisit this problem using large-scale DMRG~\cite{PhysRevLett.69.2863, PhysRevB.48.10345, SCHOLLWOCK201196} simulations on cylindrical geometries. To more reliably detect the possible weak AFM order, we apply staggered pinning fields at the open boundaries of the cylinders~\cite{PhysRevLett.99.127004}. By explicitly breaking the SU(2) symmetry, we can directly probe the bulk staggered magnetization. For each finite geometry, we perform controlled extrapolations with truncation error, followed by finite-size scaling to the thermodynamic limit. Following Ref.~\cite{PhysRevLett.99.127004}, we compare the finite-size scaling of the results with different aspect ratios and determine the staggered magnetization in the thermodynamic limit as the converged value. Our results demonstrate that the staggered magnetization remains finite throughout the previously disputed region $0.78 \lesssim J_{1}/J_{2} \leq 0.81$. These findings, together with the previously established consensus placing the PVBS phase boundary at $J_{1}/J_{2} \approx 0.78$~\cite{PhysRevX.9.041037, FRG2022, PhysRevB.105.L060409, Anders2022CPL, PhysRevB.107.L220408, chen2025spinexcitationsshastrysutherlandmodel, Viteritti2025, Qian2025, Corboz2025}, provide evidence for a direct transition between the PVBS and N\'eel AFM phases and rule out the existence of an intermediate QSL phase in the Shastry-Sutherland model.


{\em Method --}
\label{method}
DMRG has established itself as arguably the most powerful numerical method for studying one-dimensional and quasi-one-dimensional quantum many-body systems. Its efficiency originates from the underlying matrix product state (MPS) variational ansatz~\cite{PhysRevLett.75.3537}, which provides a compact representation of area-law entangled states for one-dimensional systems~\cite{SCHOLLWOCK201196, RevModPhys.82.277,RevModPhys.93.045003, xiang2023density}, thereby capturing the essential entanglement structure of the system at a relatively low computational cost. 
In DMRG, the bond dimension $D$ governs the maximum entanglement that can be captured by the state, directly determining the numerical precision of the DMRG results. DMRG can also accurately solve the ground state of spin models on two-dimensional cylinders with width around $10$ by pushing the bond dimension to very large values~\cite{PhysRevLett.113.027201, PhysRevLett.121.107202, PhysRevB.109.L161103, Huang_2024, PhysRevLett.134.196702}. In this work, we exploit the U(1) symmetry associated with total $S^z$ conservation and push the bond dimension up to as large as $D=30000$ for the widest systems ($L_{y}=12$), enabling us to suppress the truncation error to the order of $10^{-6}$ and perform reliable extrapolations for each finite-size system.

Fig.~\ref{plaquette}(a) illustrates the Shastry-Sutherland model studied in this work on cylinders, where we impose open boundary conditions (OBC) in the $x$ direction and periodic boundary conditions (PBC) in the $y$ direction. Throughout this work, we set $J_2=1$ and express all energies and pinning field strengths in units of $J_2$. We define the aspect ratio $\alpha = L_{x}/L_{y}$. To systematically approach the thermodynamic limit while mitigating finite-size effects, we mainly perform calculations on cylinders with aspect ratios $\alpha \in \{1.5, 2, 3\}$, with the cylinder width $L_y$ reaching up to 12, and check the convergence of the results with different aspect ratios~\cite{PhysRevLett.99.127004, Stoudenmire_2012, Huang_2024}. At the representative point $J_1/J_2=0.8$, we also study the square-like geometry with $\alpha=1$ as an additional check of geometry-dependent finite-size effects.

\begin{figure}[t]
    \includegraphics[width=85mm]{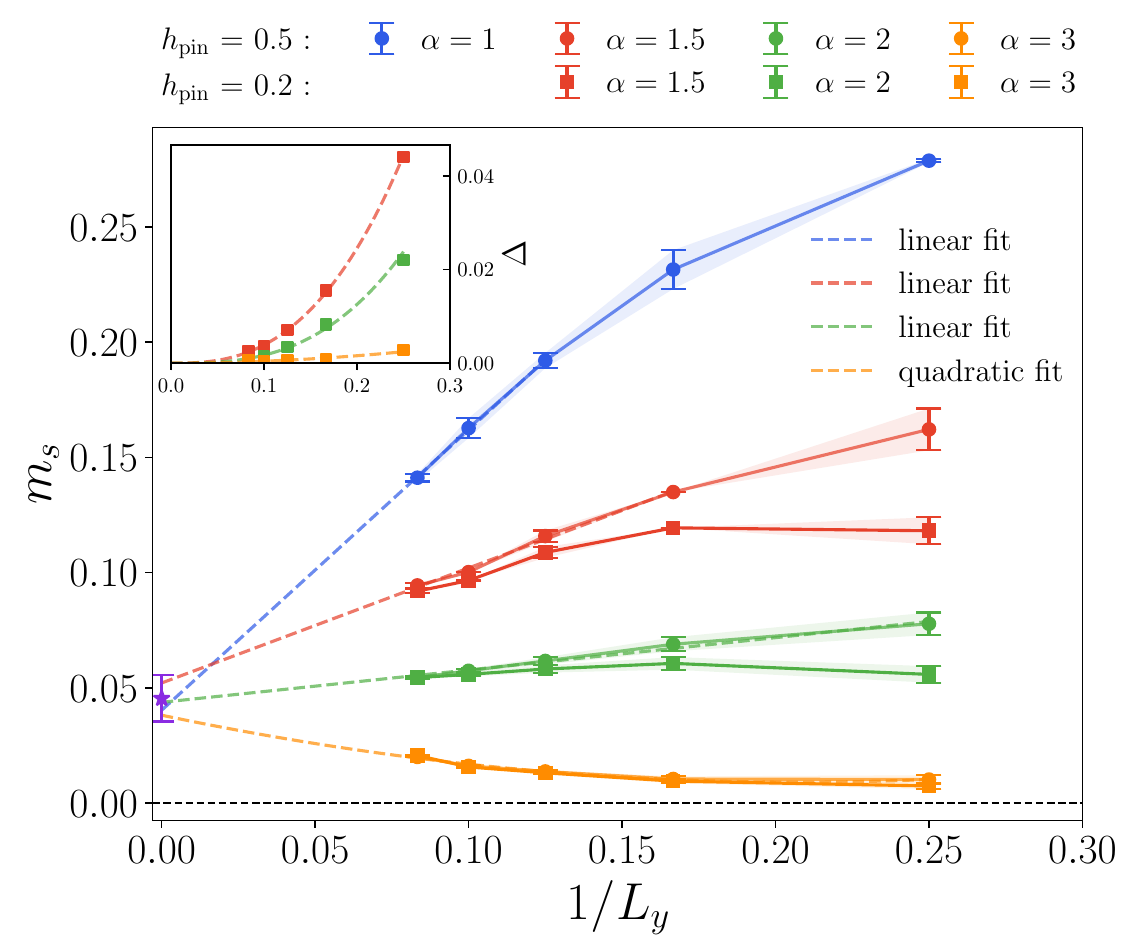}
    \caption{Finite-size scaling of the staggered magnetization $m_{s}$ at $J_{1}/J_{2}=0.8$ with two pinning field strengths, $h_{\mathrm{pin}}=0.5$ and $0.2$. Values extrapolated to zero DMRG truncation error are shown (see Figs.~S3--S5 in the Supplemental Materials~\cite{SM} for the corresponding extrapolations). Data for systems with different aspect ratios $\alpha$ are extrapolated to the thermodynamic limit ($1/L_{y} \to 0$). Inset: The difference in $m_{s}$ between the two pinning fields, denoted as $\Delta$, as a function of $1/L_{y}$. We can see $\Delta$ vanishes as $1/L_{y} \to 0$, demonstrating that the extrapolated staggered magnetization is independent of the pinning field strength. The converged finite extrapolated value of $m_s$ for systems with different aspect ratios indicates long-range N\'eel AFM order at this coupling ratio.}
    \label{pin_check}
\end{figure}

To probe the magnetic order of the system, we follow the strategy in Ref.~\cite{PhysRevLett.99.127004}. We apply staggered pinning fields along the $z$ direction with strength $h_{\mathrm{pin}}$ at the open boundaries to explicitly break the SU(2) symmetry. We set $h_{\mathrm{pin}} = 0.5$ in this work. For the $J_1/J_2=0.8$, our results show that $h_{\mathrm{pin}} = 0.2$ results agree with the $h_{\mathrm{pin}} = 0.5$ results in the thermodynamic limit, indicating our conclusion is independent of the strength of the pinning fields. The application of boundary pinning fields enables a direct measurement of the staggered magnetization $m_{s}$ in the bulk. Specifically, we define $m_s$ as
\begin{equation}
    m_s = \frac{1}{N_c} \sum_{i \in \text{central}} \eta_i \langle S^z_i \rangle,
    \label{ms_definition}
\end{equation}
where $\eta_{i} = \pm 1$ denotes the phase factor of the expected staggered pattern, and $N_{c}$ is the number of sites included in the average. The summation is taken over the central region of the cylinder to reduce boundary-induced artifacts: for cylinders with even length $L_{x}$, we consider the two central columns, as illustrated by the red dashed box in Fig.~\ref{plaquette}(a), whereas for cylinders with odd $L_{x}$, we consider only the single central column. For cylinders with even length, the onsite staggered magnetizations in the central region generally take different values on the two sublattices, denoted as $A$ and $B$, which are associated with the two orthogonal orientations of the dimer bonds in the Shastry-Sutherland lattice. However, they should converge to the same value in the thermodynamic limit. 

\begin{figure}
    \includegraphics[width=85mm]{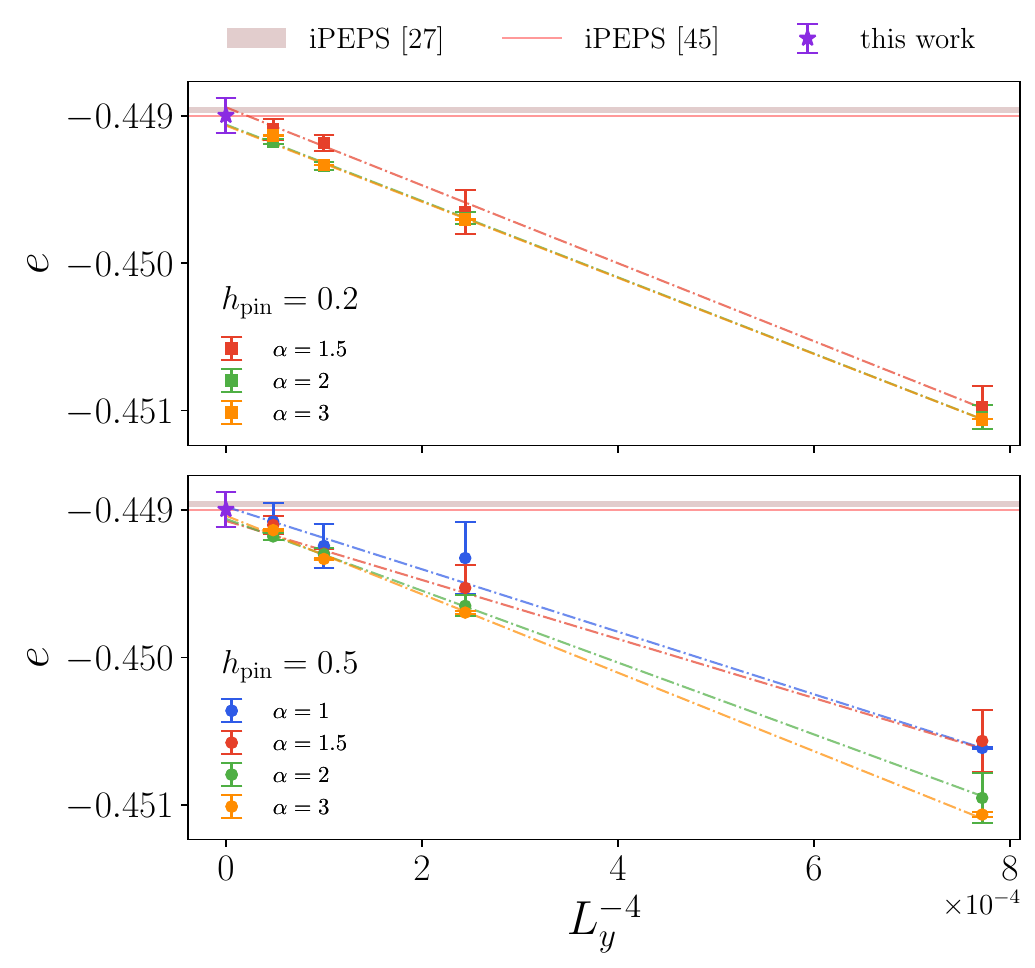}
    \caption{Finite-size scaling of the ground state energy density $e$ for the Shastry-Sutherland model at $J_{1}/J_{2}=0.8$. The bulk energies extrapolated to zero DMRG truncation error (see the Supplemental Materials~\cite{SM} for details) for $L_{y} = 6, 8, 10,$ and $12$ with different aspect ratios $\alpha$ are shown. The upper and lower panels correspond to $h_{\mathrm{pin}}=0.2$ and $0.5$, respectively. The resulting thermodynamic limit estimate, $e_{\infty}=-0.4490(1)$ (the purple star), is consistent with the previous iPEPS result~\cite{Corboz2025,SingleLayer_TN}.}
    \label{Energy_Benchmark}
\end{figure}

To account for this finite-size sublattice splitting, we perform independent linear extrapolations of the staggered magnetizations averaged separately over the $A$ and $B$ sites within the central region, denoted as $m_s^{A}$ and $m_s^{B}$, as functions of the truncation error $\varepsilon$ for each geometry. The extrapolated values in the $\varepsilon\to 0$ limit are subsequently averaged to yield the staggered magnetization for each finite geometry,
\begin{equation}
    m_s = \frac{m_s^{A} + m_s^{B}}{2}.
\end{equation}
The difference between $m_{s}^{A}$ and $m_{s}^{B}$ provides a measure of the finite-size splitting associated with the two inequivalent sublattices and is reflected in the error bars of the subsequent finite-size scaling analysis. The truncation-error extrapolations are shown in the Supplemental Materials~\cite{SM}. Finally, the resulting values of $m_s$ are extrapolated as a function of the inverse cylinder width $1/L_y$, to obtain the staggered magnetization in the thermodynamic limit.


\begin{figure*}[t]
    \includegraphics[width=175mm]{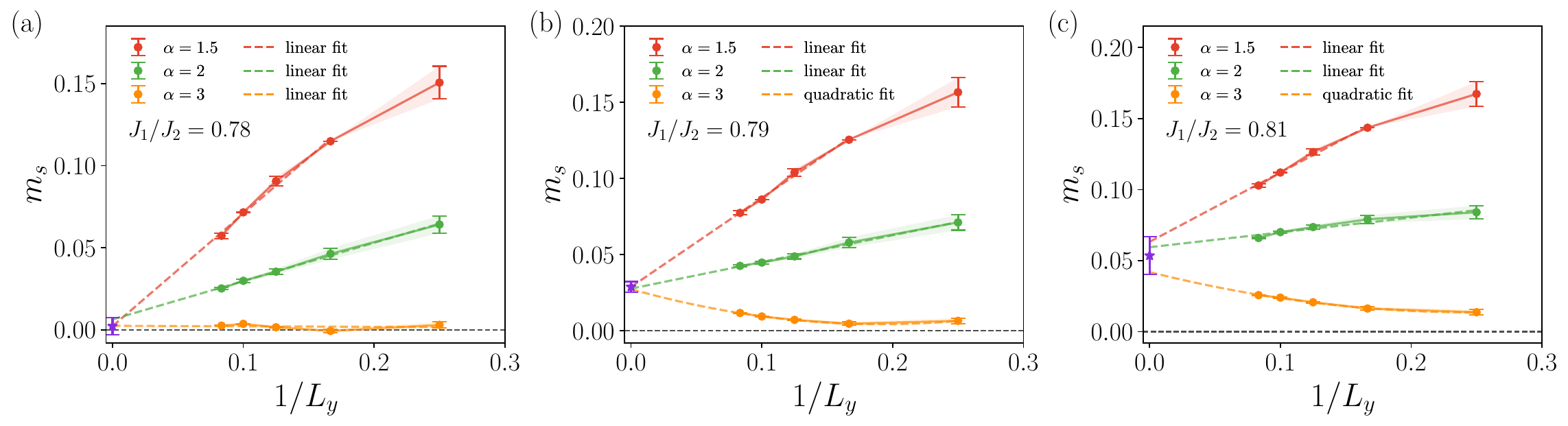}
    \caption{Finite-size scaling of the staggered magnetization $m_s$ extrapolated to zero DMRG truncation error (see the Supplemental Materials~\cite{SM} for details) for different coupling ratios: (a) $J_{1}/J_{2}=0.78$, (b) $0.79$, and (c) $0.81$. All data are obtained with a pinning field strength of $h_{\mathrm{pin}} = 0.5$. As in Fig.~\ref{pin_check}, $m_s$ converges to the same value in the thermodynamic limit for systems with different aspect ratios $\alpha$.}
    \label{Ms_center_SSM_combined}
\end{figure*}

{\em Results --}
We first investigate systems with $J_1/J_2 = 0.8$, a representative point deep inside the intermediate QSL regime proposed in several previous studies~\cite{FRG2022, PhysRevB.105.L060409, Viteritti2025, Corboz2025}. To systematically approach the thermodynamic limit while mitigating finite-size effects, we perform DMRG simulations on cylinders with four distinct aspect ratios: $\alpha \in \{ 1, 1.5, 2, 3\}$. As shown in Fig.~\ref{pin_check}, the staggered magnetization exhibits different finite-size effects as expected \cite{PhysRevLett.99.127004, Stoudenmire_2012}. For aspect ratios $\alpha = 1$ and $1.5$, $m_s$ decreases with the increase of the system size, while systems with aspect ratio $\alpha = 3$ show opposite behavior. Mild finite-size effects are found for aspect ratio $\alpha = 2$. A simple finite-size scaling of the staggered magnetization $m_s$ with a boundary pinning field $h_{\mathrm{pin}}=0.5$ yields a consistent nonzero extrapolated value across all four geometries, with $m_s = 0.045(10)$ in the thermodynamic limit. The quoted uncertainty is chosen conservatively to encompass the spread among different aspect ratios, thereby accounting for geometry-dependent effect. These results indicate that $J_{1}/J_{2} = 0.8$ lies in the N\'eel AFM phase.

We also check the effect of the strength of the pinning fields on the extrapolated staggered magnetization. As shown in the inset of Fig.~\ref{pin_check}, the difference of the values of staggered magnetization for $h_{\mathrm{pin}}=0.2$ and $h_{\mathrm{pin}}=0.5$ vanishes in the thermodynamic limit.

We also calculate the ground state energy density $e$ at $J_{1}/J_{2}=0.8$. To reduce boundary effect, we calculate local site energies from the corresponding bond energies and average them over the two central columns for even $L_x$ and the three central columns for odd $L_x$. For each finite geometry, the resulting average energy $e$ is extrapolated to zero DMRG truncation error, with details in the Supplemental Materials~\cite{SM}. The finite-size energies for $L_{y} = 6, 8, 10,$ and $12$ are then extrapolated using the power-law form $e(L_y)=e_{\infty}+A L_y^{-4}$, with the exponent $4$ chosen to maximize the coefficient of determination $R^2$ among the tested fits. As shown in Fig.~\ref{Energy_Benchmark}, the extrapolations for $h_{\mathrm{pin}}=0.2$ and $0.5$ across multiple aspect ratios yield mutually consistent estimates, giving $e_{\infty}=-0.4490(1)$, in agreement with the recent iPEPS result $e_{\infty}=-0.44896(2)$~\cite{Corboz2025} and $e_{\infty}=-0.4490$ in ~\cite{SingleLayer_TN}.


After establishing that $J_{1}/J_{2} = 0.8$ lies in the N\'eel AFM phase, we scan the interval $0.78 \le J_1/J_2 \le 0.81$ to determine the N\'eel AFM phase boundary on the small $J_{1}/J_{2}$ side. For these calculations, we use the same pinning field strength of $h_{\mathrm{pin}} = 0.5$. The systematic finite-size scaling for the specific values of $J_1/J_2 = 0.78, 0.79,$ and $0.81$ are presented in Fig.~\ref{Ms_center_SSM_combined}, showing the convergence of the staggered magnetization across multiple aspect ratios. The resulting extrapolated values of $m_s$ are summarized as a function of the coupling ratio $J_{1}/J_{2}$ in the magnetic phase diagram shown in Fig.~\ref{plaquette}(b). As captured in this phase diagram, $m_s$ exhibits a smooth and monotonic increase across the investigated interval. 

At the lower bound of our investigated range, $J_1/J_2 = 0.78$, we obtain an extrapolated magnetization of $m_s = 0.0023(53)$, suggesting this point lies at the boundary of the N\'eel AFM phase. We notice that at $J_1/J_2 = 0.78$ for the $6 \times 18$ cylinder, one of the sublattice staggered magnetization becomes slightly negative because the central region no longer display the expected perfect N\'eel staggered pattern, consistent with the strong suppression of N\'eel AFM order at this coupling. Previous studies have established a broad consensus that the PVBS phase terminates also at $J_1/J_2 \approx 0.78$~\cite{PhysRevX.9.041037, FRG2022, PhysRevB.105.L060409, Anders2022CPL, PhysRevB.107.L220408, chen2025spinexcitationsshastrysutherlandmodel, Viteritti2025, Qian2025, Corboz2025}. So within the resolution of our calculations, we find no evidence for a stable intermediate QSL phase, supporting instead a direct transition between the PVBS and N\'eel AFM phases.

{\em Conclusion --}
In this work, we determine the precise boundary of the N\'eel AFM phase of the Shastry-Sutherland model using large-scale DMRG calculations. By employing staggered boundary pinning fields, we directly probe the bulk magnetic order parameter in the controversial region of $0.78 \le J_{1}/J_{2} \le 0.81$. Our results demonstrate that the N\'eel AFM order remains finite at $J_{1}/J_{2} = 0.8$, a point which was believed to lie in an intermediate spin liquid phase~\cite{FRG2022, PhysRevB.105.L060409, Viteritti2025, Corboz2025}. We find that the boundary of the N\'eel AFM phase is at $J_{1}/J_{2} \approx 0.78$, where the PVBS phase terminates~\cite{PhysRevX.9.041037, FRG2022, PhysRevB.105.L060409, Anders2022CPL, PhysRevB.107.L220408, chen2025spinexcitationsshastrysutherlandmodel, Viteritti2025, Qian2025, Corboz2025}. These results provide evidence against an intermediate QSL phase in the Shastry-Sutherland model. Our findings instead support a direct transition between the PVBS and N\'eel AFM phases.


\begin{acknowledgments}
The calculation in this work is carried out with TensorKit~\cite{foot7}. The computations in this paper were run on the Siyuan-1 cluster supported by the Center for High Performance Computing at Shanghai Jiao Tong University. MQ acknowledges the support from the National Natural Science Foundation of China (Grant No. 12522406 and No. 12274290), the Innovation Program for Quantum Science and Technology (2021ZD0301902), and the National Key Research and Development Program of MOST of China (2022YFA1405400).
\end{acknowledgments}

\bibliography{main}

\clearpage

\onecolumngrid

\renewcommand{\thefootnote}{\fnsymbol{footnote}}
\setcounter{footnote}{0}

\begin{center}
{\large\bfseries
Supplemental Materials for\\
``On the precise boundary of the N\'eel antiferromagnetic phase in the Shastry-Sutherland model''\\[0.9em]
}

Rongyi Lv,$^{1}$ Xiangjian Qian,$^{1, 2}$ and Mingpu Qin$^{1, 3, }$\footnotemark \\[0.4em]

{\small
\textit{$^{1}$Key Laboratory of Artificial Structures and Quantum Control (Ministry of Education),  \\School of Physics and Astronomy, Shanghai Jiao Tong University, Shanghai 200240, China\\
$^{2}$Tsung-Dao Lee Institute, Shanghai Jiao Tong University, Shanghai 200240, China\\
$^{3}$Hefei National Laboratory, Hefei 230088, China}
}
\\
(Dated: \today)
\\[1.6em]
\end{center}

\footnotetext[1]{qinmingpu@sjtu.edu.cn}

\twocolumngrid

\setcounter{section}{0}
\setcounter{figure}{0}
\setcounter{table}{0}
\setcounter{equation}{0}

\renewcommand{\thefigure}{S\arabic{figure}}
\renewcommand{\thetable}{S\arabic{table}}
\renewcommand{\theequation}{S\arabic{equation}}

\renewcommand{\theHsection}{S\arabic{section}}
\renewcommand{\theHfigure}{S\arabic{figure}}
\renewcommand{\theHtable}{S\arabic{table}}
\renewcommand{\theHequation}{S\arabic{equation}}

\section{Truncation-error extrapolations of the staggered magnetization}
\label{extrapolation_truncation_error}
\begin{figure*}[t]
    \includegraphics[width=165mm]{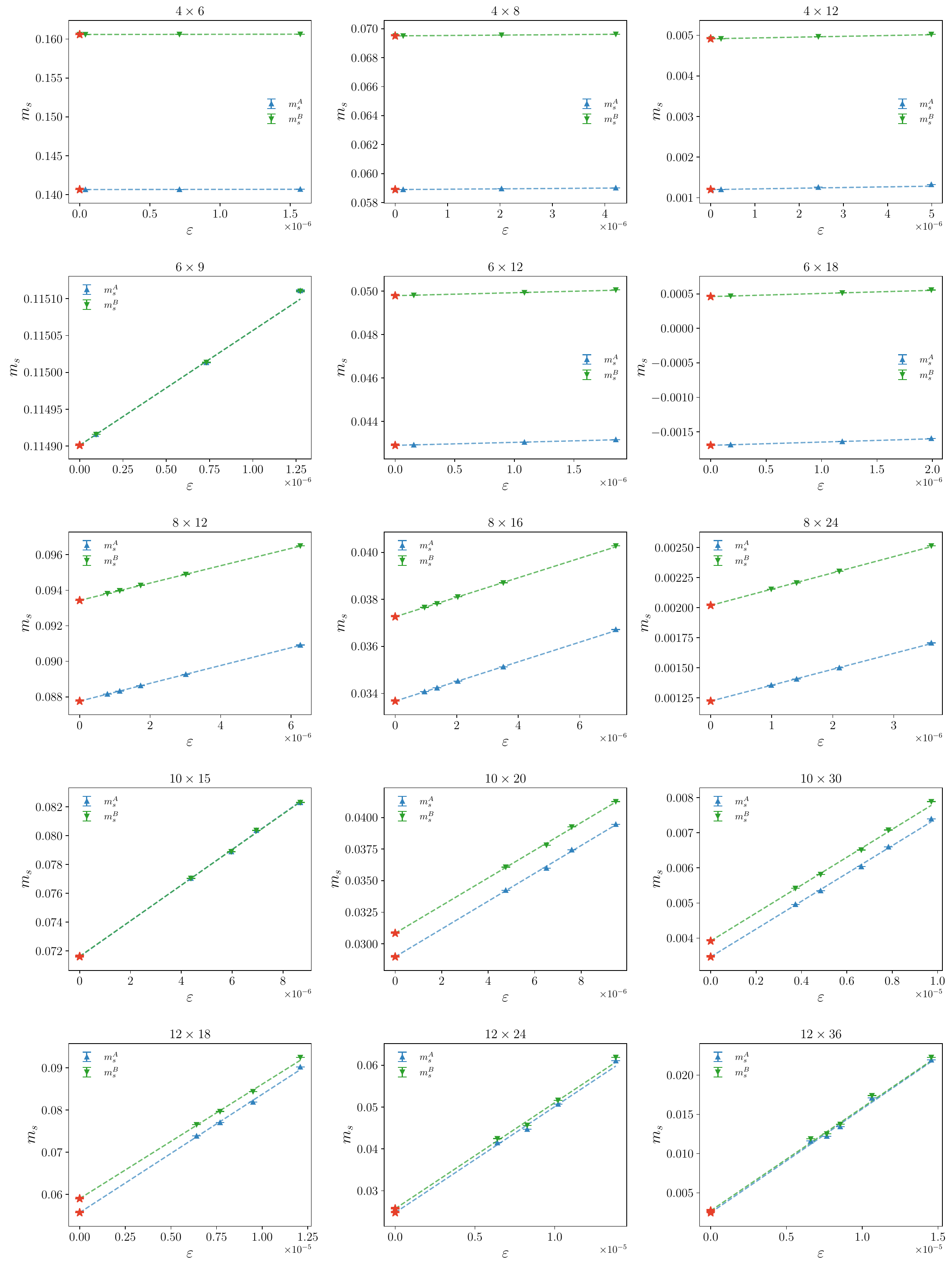}
    \caption{Extrapolations of the sublattice staggered magnetizations $m_s^{A}$ and $m_s^{B}$ as functions of the truncation error $\varepsilon$ for different system sizes $(L_y \times L_x)$ at $J_1/J_2=0.78$ with $h_{\mathrm{pin}}=0.5$.}
    \label{Combined_Grid_780_pin_05}
\end{figure*}

\begin{figure*}[t]
    \includegraphics[width=165mm]{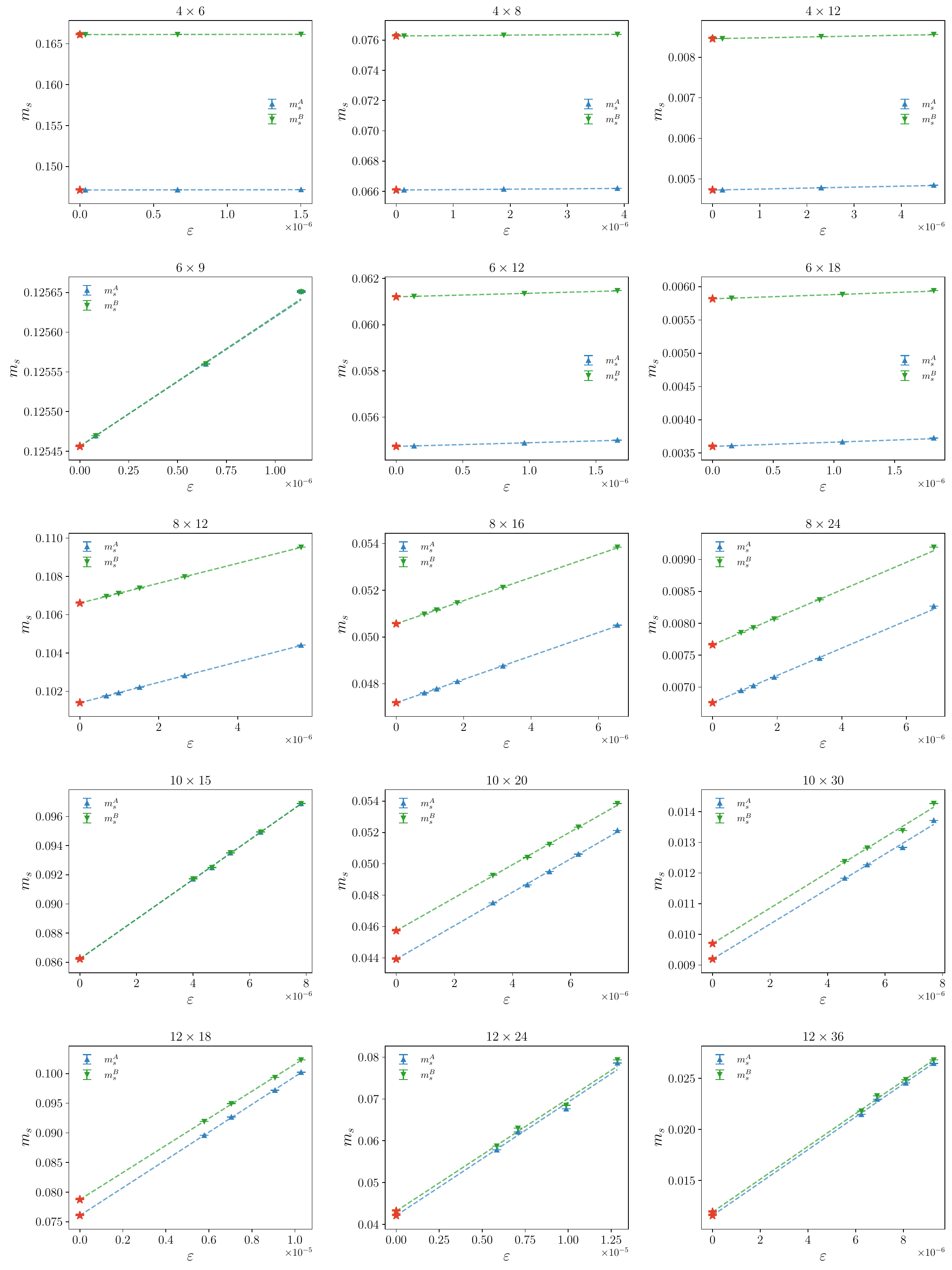}
    \caption{Extrapolations of the sublattice staggered magnetizations $m_s^{A}$ and $m_s^{B}$ as functions of the truncation error $\varepsilon$ for different system sizes $(L_y \times L_x)$ at $J_1/J_2=0.79$ with $h_{\mathrm{pin}}=0.5$.}
    \label{Combined_Grid_790_pin_05}
\end{figure*}

\begin{figure*}[t]
    \includegraphics[width=165mm]{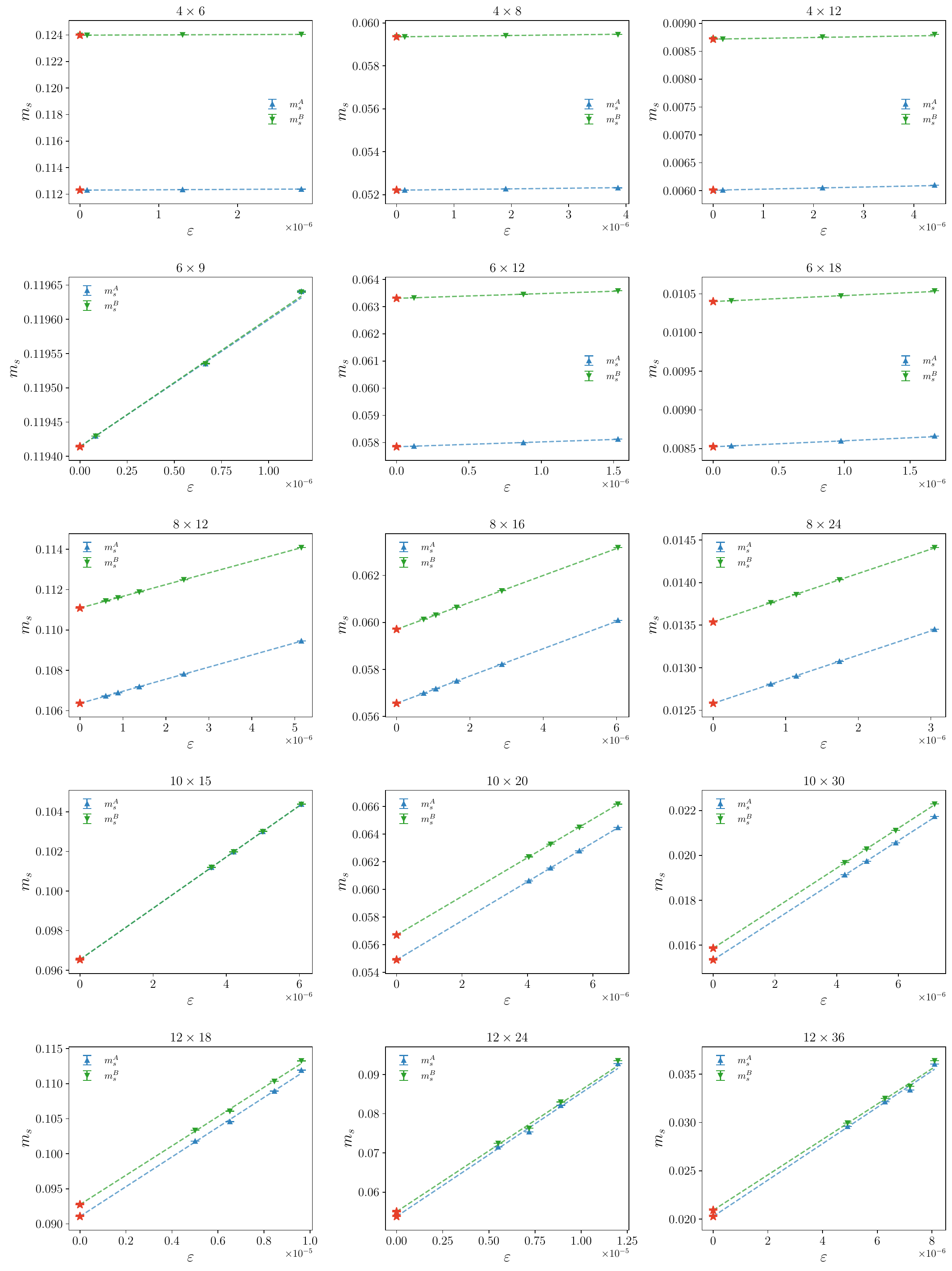}
    \caption{Extrapolations of the sublattice staggered magnetizations $m_s^{A}$ and $m_s^{B}$ as functions of the truncation error $\varepsilon$ for different system sizes $(L_y \times L_x)$ at $J_1/J_2=0.8$ with $h_{\mathrm{pin}}=0.2$.}
    \label{Combined_Grid_800_pin_02}
\end{figure*}

\begin{figure*}[t]
    \includegraphics[width=165mm]{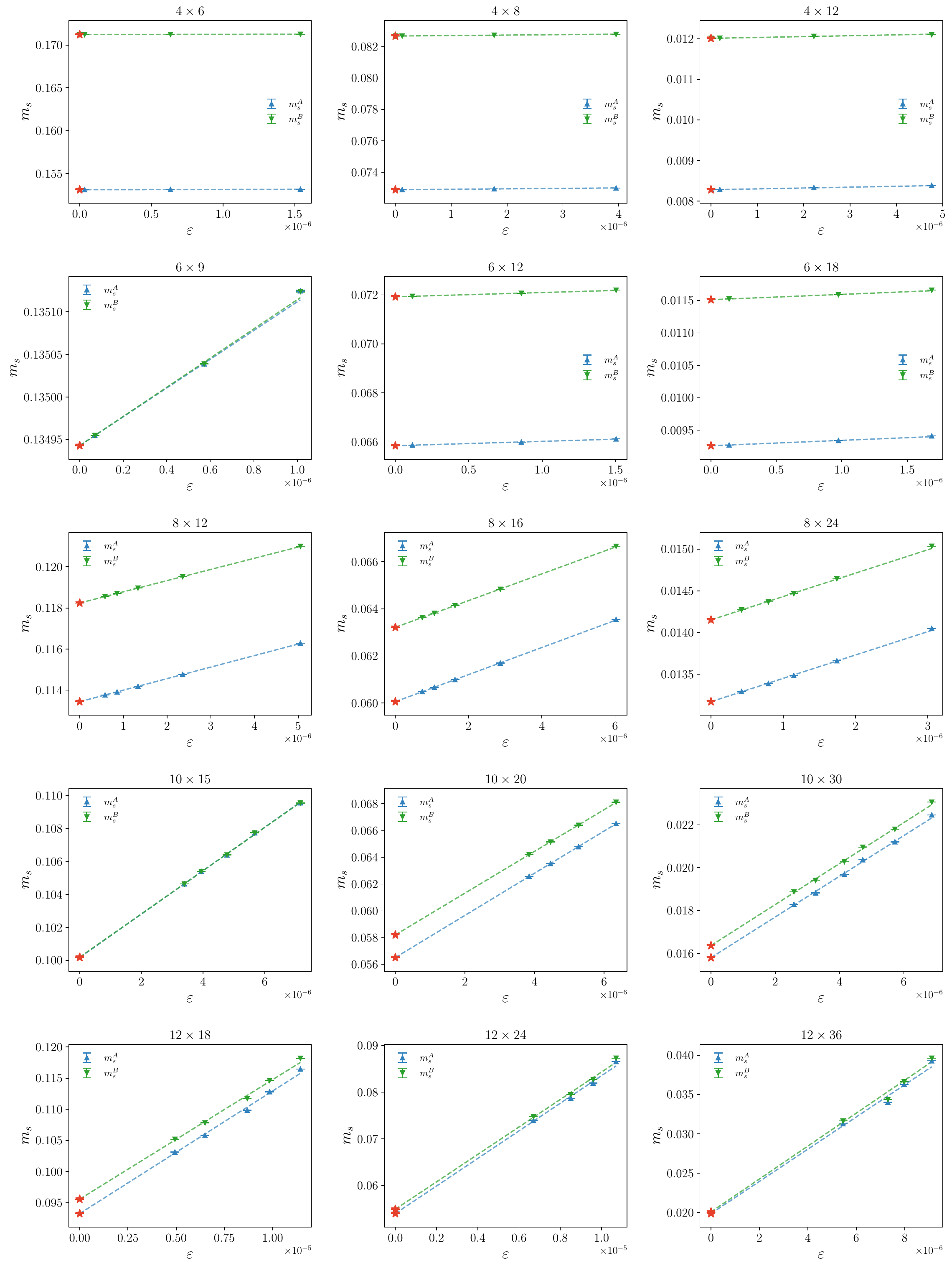}
    \caption{Extrapolations of the sublattice staggered magnetizations $m_s^{A}$ and $m_s^{B}$ as functions of the truncation error $\varepsilon$ for different system sizes $(L_y \times L_x)$ at $J_1/J_2=0.8$ with $h_{\mathrm{pin}}=0.5$.}
    \label{Combined_Grid_800_pin_05}
\end{figure*}

\begin{figure*}[t]
    \includegraphics[width=165mm]{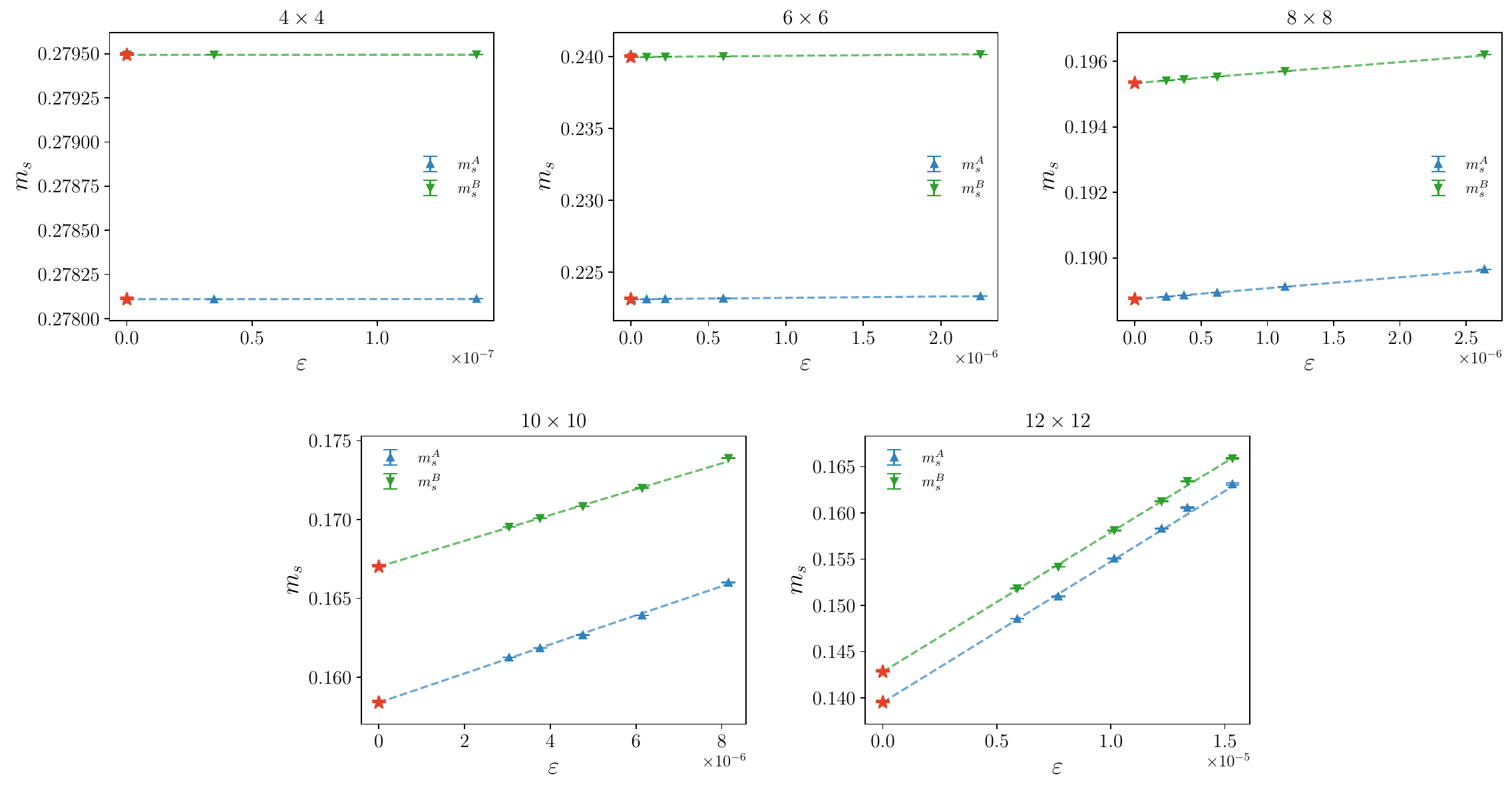}
    \caption{Extrapolations of the sublattice staggered magnetizations $m_s^{A}$ and $m_s^{B}$ as functions of the truncation error $\varepsilon$ for different system sizes $(L_y \times L_x)$ at $J_1/J_2=0.8$ with $h_{\mathrm{pin}}=0.5$.}
    \label{Combined_Grid_800_pin_05_LxL_plots}
\end{figure*}

\begin{figure*}[t]
    \includegraphics[width=165mm]{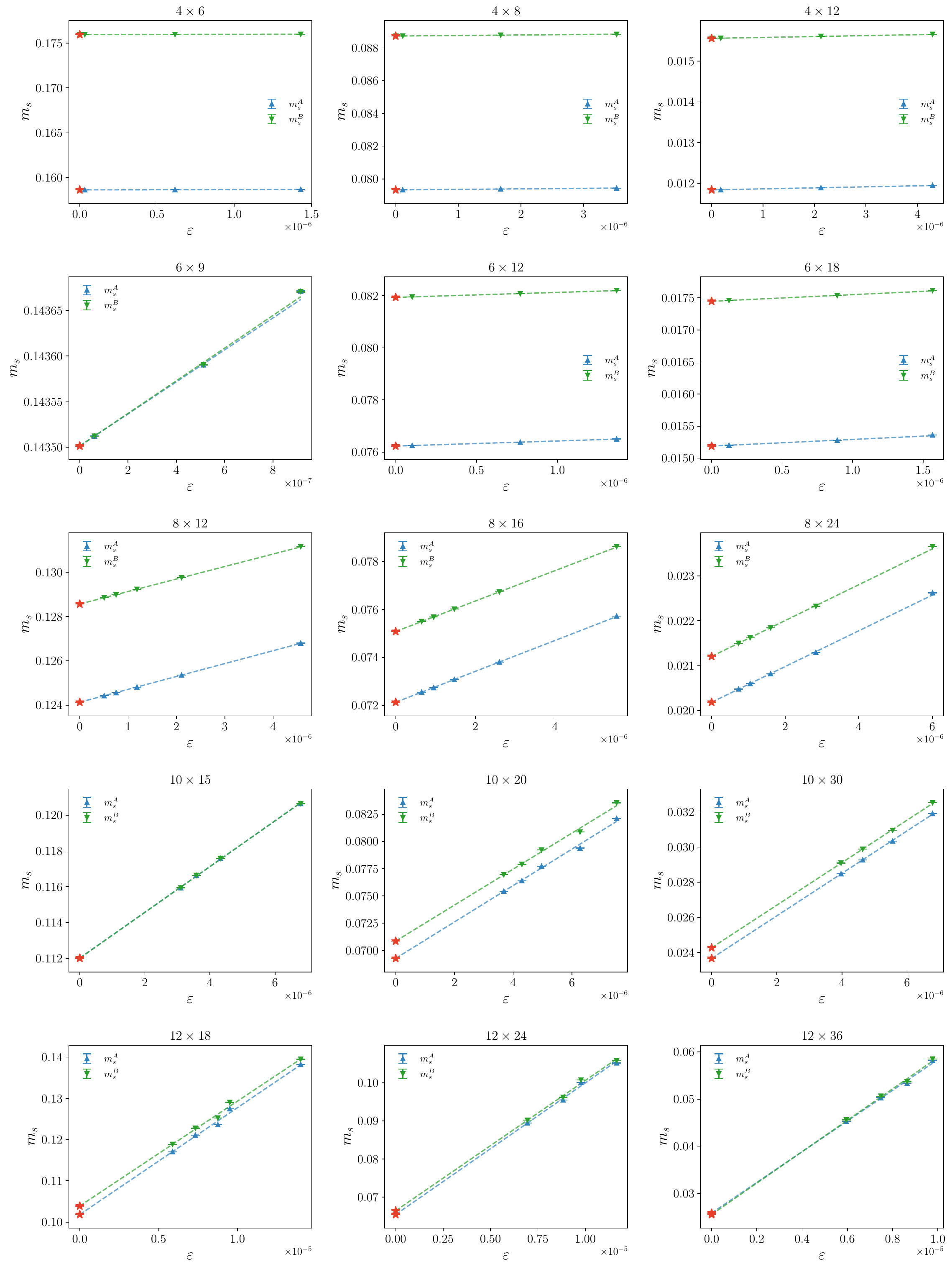}
    \caption{Extrapolations of the sublattice staggered magnetizations $m_s^{A}$ and $m_s^{B}$ as functions of the truncation error $\varepsilon$ for different system sizes $(L_y \times L_x)$ at $J_1/J_2=0.81$ with $h_{\mathrm{pin}}=0.5$.}
    \label{Combined_Grid_810_pin_05}
\end{figure*}

Following the procedure described in the main text, we perform linear extrapolations of the staggered magnetizations $m_s^A$ and $m_s^B$, averaged separately over the $A$ and $B$ sites within the central region, as functions of the truncation error $\varepsilon$ for each finite geometry. All truncation-error extrapolations used in the finite-size scaling analysis are shown in Figs.~\ref{Combined_Grid_780_pin_05}, \ref{Combined_Grid_790_pin_05}, 
\ref{Combined_Grid_800_pin_02}, \ref{Combined_Grid_800_pin_05}, \ref{Combined_Grid_800_pin_05_LxL_plots}, and~\ref{Combined_Grid_810_pin_05}.

\section{Extrapolations of the ground state energies}

\begin{figure*}
    \includegraphics[width=100mm]{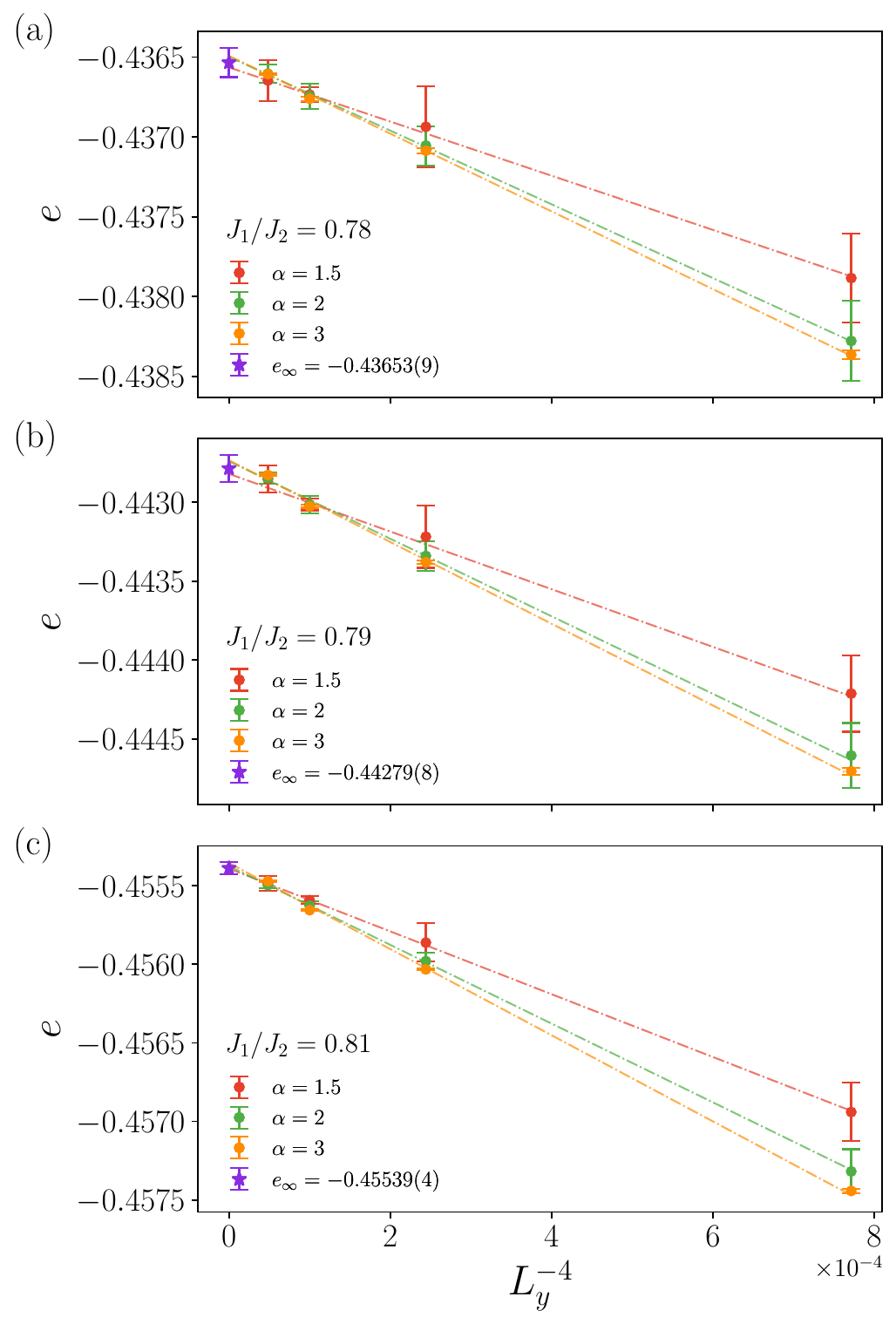}
    \caption{Finite-size scaling of the ground state energy density $e$ for the Shastry-Sutherland model at different coupling ratios: (a) $J_{1}/J_{2}=0.78$, (b) $0.79$, and (c) $0.81$. All data are obtained with a pinning field strength of $h_{\mathrm{pin}}=0.5$. The bulk energies extrapolated to zero DMRG truncation error are shown (see Fig.~\ref{Energy_Extrapolations_Combined} for the corresponding extrapolations) and are further extrapolated using $e(L_y)=e_{\infty}+A L_y^{-4}$.}
    \label{Energy_CenterBonds_Combined}
\end{figure*}

\begin{figure*}
    \includegraphics[width=55mm]{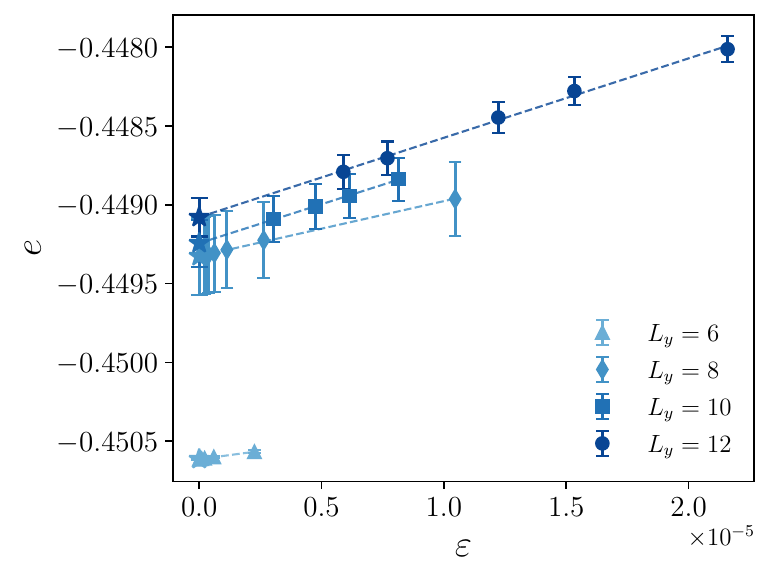}
    \caption{Extrapolations of the bulk ground state energy density $e$ as functions of the truncation error $\varepsilon$ for aspect ratio $\alpha=1$ at $J_1/J_2=0.8$ with $h_{\mathrm{pin}}=0.5$.}
    \label{Energy_Extrapolations_0.8_1.0_0.5_alpha1}
\end{figure*}

\begin{figure*}[t]
    \includegraphics[width=165mm]{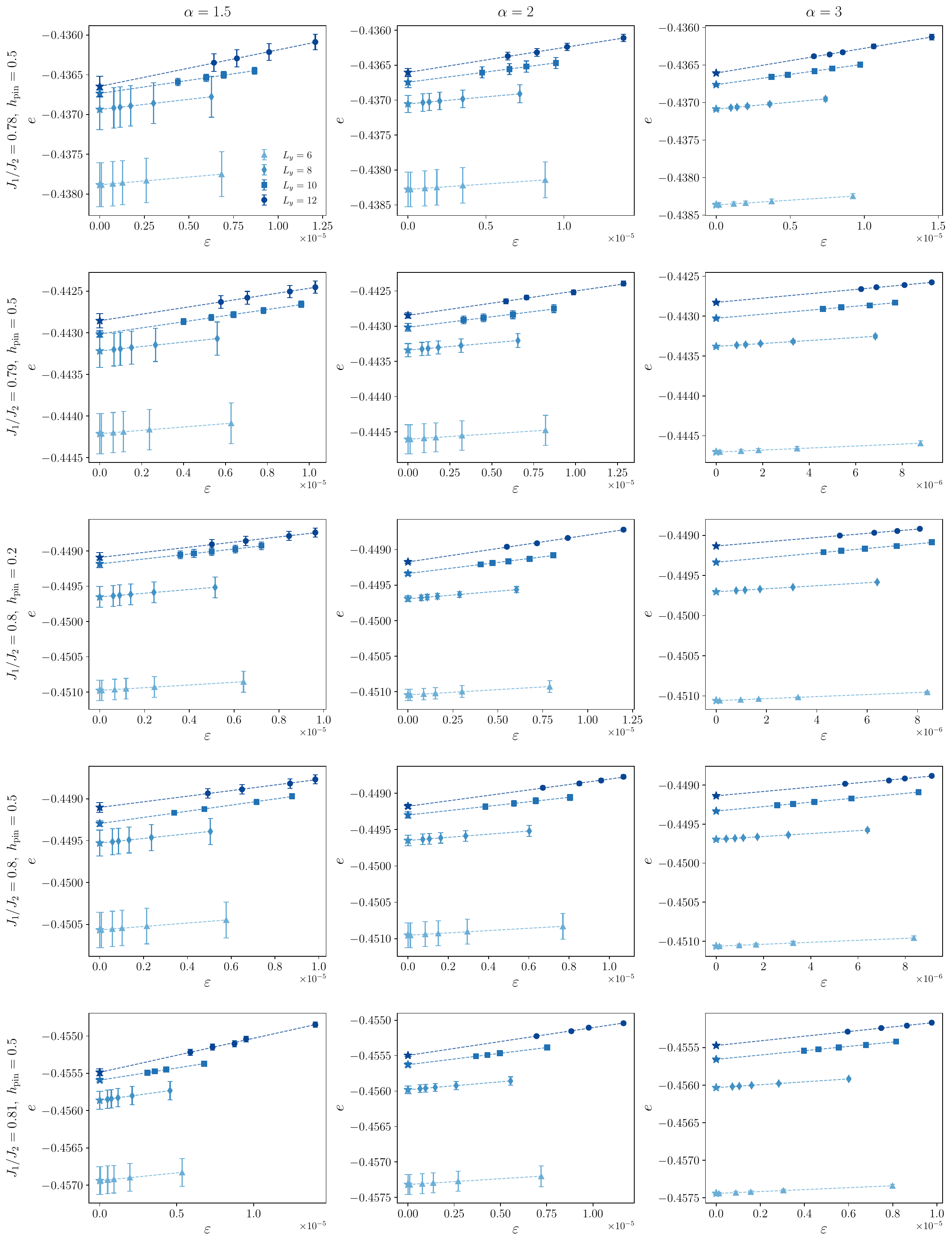}
    \caption{Extrapolations of the bulk ground state energy density $e$ as functions of the truncation error $\varepsilon$. From top to bottom, the five rows correspond to $(J_1/J_2,h_{\mathrm{pin}})=(0.78,0.5)$, $(0.79,0.5)$, $(0.8,0.2)$, $(0.8,0.5)$, and $(0.81,0.5)$, while the three columns correspond to aspect ratios $\alpha=1.5$, $2$, and $3$, respectively.}
    \label{Energy_Extrapolations_Combined}
\end{figure*}

We further determine the thermodynamic limit ground state energy densities at $J_{1}/J_{2}=0.78$, $0.79$, and $0.81$ using the same finite-size scaling procedure as in Fig.~3 of the main text. Fig.~\ref{Energy_CenterBonds_Combined} shows the finite-size scaling of the bulk energies for these coupling ratios on cylinders with different aspect ratios. Each data point is obtained by linearly extrapolating the corresponding finite-geometry bulk energy to zero DMRG truncation error. The complete set of truncation-error extrapolations of the energies used in this work is presented in Figs.~\ref{Energy_Extrapolations_0.8_1.0_0.5_alpha1} and~\ref{Energy_Extrapolations_Combined}.

\end{document}